\documentclass{PoS}

\usepackage{amsmath,amssymb,amsfonts}
\usepackage{algorithm}
\usepackage{algorithmic}
\usepackage{xspace}
\usepackage{graphicx}
\usepackage{fontenc,times,mathptmx}

\graphicspath{{./}{pics/}}

\newcommand{\CC}{\mathbb{C}}
\newcommand{\Span}[1]{\textnormal{span}\left\{#1\right\}}
\newcommand{\Colspan}[1]{\textnormal{colspan}#1}

\algsetup{
 linenosize=\small,
 linenodelimiter=.
}

\def\BlockLanczos{block Lanczos\xspace}
\def\BlockLanczosProcess{\BlockLanczos process\xspace}
\def\BlockKrylovSubspace{block Krylov subspace\xspace}
\def\BBlockKrylovSubspace{Block Krylov subspace\xspace}
\def\DSBlockCG{DSBlockCG\xspace}

\def\rhs{right hand side\xspace}
\def\rhss{right hand sides\xspace}

\title{A CG Method for Multiple Right Hand Sides and Multiple Shifts in Lattice QCD Calculations}

\ShortTitle{A CG Method for Multiple Right Hand Sides and Multiple Shifts in Lattice QCD Calculations}

\author{\speaker{Sebastian Birk} 
         \thanks{This work was supported by Deutsche Forschungsgemeinschaft through the Collaborative Research Centre SFB-TR 55 "Hadron physics from Lattice QCD"}\\
        Fachbereich C, Mathematik und Naturwissenschaften, Bergische Universit\"at Wuppertal, D-42097 Wuppertal, Germany\\
        E-mail: \email{birk@math.uni-wuppertal.de}}

\author{Andreas Frommer\\
       Fachbereich C, Mathematik und Naturwissenschaften, Bergische Universit\"at Wuppertal, D-42097 Wuppertal, Germany\\
       E-mail: \email{frommer@math.uni-wuppertal.de}}

\abstract{We consider the task of computing solutions of linear systems that only differ by a shift with the identity matrix as well as linear systems with several different \rhss{}. In the past Krylov subspace methods have been developed which exploit either the need for solutions to multiple \rhss{} (e.g.\ deflation type methods and block methods) or multiple shifts (e.g.\ shifted CG) with some success.  In this paper we present a \BlockKrylovSubspace{} method which, based on a \BlockLanczosProcess{}, exploits both features---shifts and multiple \rhss{}---at once. Such situations arise, for example, in lattice QCD simulations within the Rational Hybrid Monte Carlo algorithm. We give numerical evidence that our method is superior to applying other iterative methods to each of the systems individually as well as, in some cases, to shifted or \BlockKrylovSubspace{} methods.}

\FullConference{XXIX International Symposium on Lattice Field Theory \\
		 July 10-16, 2011\\
		 Squaw Valley, Lake Tahoe, California}

\begin{document}

\section{Introduction}\label{sec_introduction}

In this paper we consider solving systems of linear equations of the form
\begin{align}\label{eqn_Systems}
  (\sigma_jI+A)x_{i,j} &= b_i
\end{align}
where $\sigma_jI+A\in\CC^{n\times n}$ is a hermitian positive definite (hpd) matrix for every shift $\sigma_j\in\CC$, $x_{i,j}, b_i\in\CC^{n}$, $i=1,\ldots,m$ and $j=1,\ldots,s$.

Systems like these arise naturally in lattice QCD, where the $b_i$ represent multiple source terms, and $A$ is a discrete version of the Dirac operator like, e.g.\, the Wilson fermion matrix. For example the Rational Hybrid Monte Carlo algorithm \cite{Kennedy2006} needs to solve $A^{\nu}x_i=b_i$ with $\nu\in (-1,1)$ for multiple \rhss{} (rhs) $b_i$. The matrix power $A^{\nu}$ therein is computed using a rational approximation which can be represented as a partial fraction expansion thus giving rise to multiple shifted systems. Another application is the computation of symplectic integrators that require the evaluation of $A^{\nu}BA^{\nu}x_i=b_i$ for multiple random \rhss{}. In that case the inner application of $A^{\nu}$ is approximated via a partial fraction expansion that generates multiple \rhss{} for the outer application even if there was only one \rhs{} $b_1$.

By arranging the $m$ \rhss{} and the corresponding solutions in the matrices
\[B=\left[b_1|\cdots|b_m\right]  \textnormal{ and } X_j=\left[x_{1,j}|\cdots|x_{m,j}\right] \]
we can rewrite the systems (\ref{eqn_Systems}) as
\begin{align}\label{eqn_BlockSystems}
  (\sigma_jI+A)X_{j} &= B.
\end{align}

In the past Krylov subspace methods have been developed which exploit the composition of multiple \rhss{} to blocks and solve each of the $j$ systems in \eqref{eqn_BlockSystems} in one go \cite{O'Leary1980,Dubrulle2001}. It was realised that in these so called block methods deflation, i.e.\ removing those vectors that become linearly dependent, is important in order to guarantee convergence\cite{Gutknecht2005}. Other algorithms make use of multiple shifts for systems with a single \rhs{} and compute solutions for every shifted system by spanning the Krylov subspace just once \cite{FrommerMaass1999}. For non-hermitian matrices there even exists a method that combines multiple \rhss{}--including deflation--and shifted systems \cite{FreundFrommerFiebach1997}. All these methods improved the computational costs for solving the kind of problem they are focused on. 

We here present a new, deflated shifted block CG (DSBlockCG) method that merges every aspect: the block idea, the idea of computing solutions for the shifted systems alongside one seed system and the focus on hpd matrices. It is based on a \BlockLanczosProcess{} which is the restriction to the hermitian case of the non-symmetric \BlockLanczosProcess{} from \cite{AliagaBoleyFreundHernandez2000}. This process is capable of handling multiple starting vectors and includes deflation.

\section{\BBlockKrylovSubspace{} methods}\label{sec_BlockMethods}
For ease of presentation we first disregard the shifts, i.e.\ we deal with systems
\[
Ax_i = b_i, i=1,\ldots,m.
\]
We define the \textit{$k$-th block Krylov subspace} with respect to $A$ and $B=[b_1|\cdots|b_m]$ as
\begin{equation}\label{eqn_BlockKrylovSubspace}
  \begin{aligned}
    K_k(A,B) &=\phantom{:} \Span{b_1,\ldots,b_m,Ab_1,\ldots,Ab_m,A^2b_1,\ldots,A^{k-1}b_m}\\
             &=: \Colspan{\left[\ B \ | \ AB \ | \ A^2B \ | \ \cdots \ | \ A^{k-1}B \ \right]}.
  \end{aligned}
\end{equation}
Clearly, all the Krylov subspaces $K_k(A,b_i) := \Span{b_i, Ab_i, \ldots, A^{k-1}b_i}$ are contained in $K_k(A,B)$. While the dimension of each space $K_k(A,b_i)$ is $k$ (unless we have reached an invariant subspace), the dimension of the block subspace can be smaller than $mk$. This is due to the fact that some of the subspaces $K_k(A,b_i)$ can have non-trivial intersections even when all the $b_i$ are linearly independent. 

The block conjugate gradient methods in~\cite{O'Leary1980} and \cite{Dubrulle2001} create block iterates $X^{(k)}=[x_1^{(k)}|\ldots|x_m^{(k)}]$ with $x_i^{(k)}\in K_k(A,B)$ and advance in each step from $K_k(A,B)$ to $K_{k+1}(A,B)$. The iterates $X^{(k)}$ are obtained such that they satisfy the Galerkin condition 
\[
X^{(k)} \in K_k(A,B) \mbox{ and } R^{(k)}=B-AX^{(k)}\perp K_k(A,B),
\]
which in the non-block case reduces to the classical variational characterisation of the CG iterates. Let $V^{(k)}$ denote a
matrix whose columns form a basis of $K_k(A,B)$. Then the Galerkin condition is equivalent to
\begin{equation} \label{eq_Galerkin2}
X^{(k)} = V^{(k)} \cdot \left((V^{(k)})^H AV^{(k)}\right)^{-1} \cdot (V^{(k)})^H B.
\end{equation}

\section{The deflated shifted block CG method}\label{sec_DSBlockCG}

Based on the \BlockLanczosProcess{}\cite{AliagaBoleyFreundHernandez2000} we can now explain the \DSBlockCG{} method. We start by considering the unshifted block system $AX=B$. Handling additional shifts will be addressed at the end of this section. For details on the derivation and implementation of the algorithm we refer to our upcoming paper \cite{BirkFrommer2011DSBlockCG}.

The \BlockLanczosProcess{} generates matrices $T^{(k)}$ and $V^{(k)}$. The orthonormal columns of the latter form a basis of a $k$-dimensional subspace which we will call the \textit{$k$-th deflated block Krylov subspace} $K^{\textrm{defl}}_k(A,B)$. Additionally the relation
\begin{equation} \label{eq_Tdef}
  T^{(k)} = (V^{(k)})^HAV^{(k)}.
\end{equation}
holds and $T^{(k)}$ is a hermitian, banded matrix with semi-bandwidth $m$. Note, that the dimension $k$ of $K^{\textrm{defl}}_k(A,B)$ differs from the dimension of the non-deflated subspace $K_k(A,B)$ in \eqref{eqn_BlockKrylovSubspace} which is at most $mk$, but might be less if some vectors spanning $K_k(A,B)$ are linearly dependent. As soon as such deflation occurs, the bandwidth of the trailing right lower submatrix of $T^{(k)}$ decreases accordingly.

A deflated \BlockKrylovSubspace method generates iterates $X^{(k)}=[x^{(k)}_1|\dots|x^{(k)}_m]$, s.t.\ $x^{(k)}_j\in K^{\textrm{defl}}_k(A,B)$. Building upon \eqref{eq_Galerkin2}, but using the orthogonal basis of the deflated block Krylov 
subspace and \eqref{eq_Tdef}, we obtain the iterates $X^{(k)}$ as 
\begin{equation}\label{eqn_ApproxIterate}
  X^{(k)} := V^{(k)}(T^{(k)})^{-1}(V^{(k)})^HB.
\end{equation}

In order to obtain a feasible iterative method, cheap updates for the iterates have to be obtained. By computing a root-free Cholesky decomposition $L^{(k)}D^{(k)}(L^{(k)})^H$ of the matrix $T^{(k)}$ we can update the iterates via
\begin{eqnarray*}
  X^{(k)} &=& V^{(k)}(L^{(k)}D^{(k)}(L^{(k)})^H)^{-1}(V^{(k)})^HB\\
          &=& X^{(k-1)} + \frac{1}{d^{(k)}}p^{(k)}u^{(k)}
\end{eqnarray*}
where $d^{(k)}$ is the $(k,k)$-entry in $D^{(k)}$, $p^{(k)}\in\CC^{n}$ and $u^{(k)}\in\CC^{1\times m}$. The important property for the feasibility of our method is that the Cholesky decomposition as well as $p^{(k)}$ and $u^{(k)}$ can be updated with short recurrences of length $m$ or even less if deflation occurred.

In order to be able to stop the iteration we should be able to compute the norms of the residuals $r_j^{(k)}=b_j-Ax_j^{(k)}, 1\leq j\leq m$. Fortunately, although the residuals are not directly available, their norms can be computed at very low additional cost. The iterates $X^{(k)}$ are generated in such a way, that the residuals $R^{(k)}$ are orthogonal to the Lanczos vectors $v^{(j)}$ for $j\leq {i-m}$. They may thus be written as 
\[R^{(k)}=W^{(k)}C^{(k)}\]
where $C^{(k)}\in\CC^{m\times m}$ and $W^{(k)} = [v^{(k)}|\ldots|v^{(i+m-1)}] \in \CC^{n\times m}$. Thus, if we are only interested in the norm of each of the residuals we can just compute the norm of the columns of $C^{(k)}\in\CC^{m\times m}$, because $W^{(k)}$ has orthonormal columns. Indeed, there is a computationally cheap way to compute $C^{(k)}$ using only the matrix $T^{(k)}$ and its Cholesky decomposition. If deflation occurred the number of columns of $W^{(k)}$ and rows of $C^{(k)}$ even decreases.

The block method so far can be extended to handle shifted systems and multiple \rhss{} at the same time. For ease of notation we focus on a situation where we have just one additional shift $\sigma$ and use the notation $A_\sigma:=\sigma I+A$. Matrices and vectors belonging to the shifted system will also be noted by the index $\sigma$. Krylov subspaces are shift invariant, i.e.\ $K_k(A,b) = K_k(A_\sigma,b)$ for all $k$, see \cite{paige1995approximate}.
Moreover, starting with the initial vector $b$, the Lanczos process produces exactly the same vectors, whether we take $A$ or $A_\sigma$. This property immediately carries over to the \BlockLanczosProcess{} and to the deflated \BlockLanczosProcess{}.
We can make use of
\[(V^{(k)})^H(\sigma I_n+A)V^{(k)} = \sigma I_k + T^{(k)} =: T^{(k)}_\sigma,\]
so that we do not have to compute $V^{(k)}$ and $T^{(k)}$ for each shifted system individually. Instead we can reuse these from the unshifted system $AX=B$ and only have to compute the Cholesky decomposition $L^{(k)}_\sigma$ and $D^{(k)}_\sigma$ as well as the vectors $p^{(k)}_\sigma$ and $u^{(k)}_\sigma$ for each shift.

\section{Numerical results}\label{sec_NumericalResults}

In this section we present and discuss results of applying the algorithm from section~\ref{sec_DSBlockCG} to some QCD test problems. We were working on non-parallelised algorithms, therefore our tests were restricted to rather small lattice sizes. A Chroma implementation of the algorithm is currently under development.

In order to obtain fair time measurements we implemented all of the algorithms in \texttt{C++} using the uBLAS library \cite{boost:lib} for sparse and dense BLAS operations. We stored the Wilson Dirac operator $D$ as a sparse matrix and did not exploit any further properties like its four dimensional lattice structure or symmetries. We chose a quenched configuration with temperature $\beta=6.0$.
As compiler we used g++ in version 4.5.1. The results were produced on a Core 2 Quad Q9650 running at $3.00$ GHz. Since our programs were not parallelized they were running a single core of that machine.

We compare our algorithm to three other algorithms. The first of them is non-preconditioned conjugate gradients (CG). For CG we have to solve each of the systems 
\begin{align*}
  (\sigma_jI+A)x_{i,j} &= b_i
\end{align*}
separately. The second algorithm is the shifted conjugate gradients algorithm from \cite{FrommerMaass1999} referred to as shiftedCG. It has to be applied $m$ times, once for each \rhs{} $b_i$. For each \rhs{} all $s$ shifted systems are solved at the same time. The third and last competing algorithm is BCGrQ from \cite{Dubrulle2001}. For each shift, this block algorithm solves all the systems belonging to the various \rhss{} in one go. 

\begin{figure}[ht]
  \centerline{\includegraphics[width=0.48\textwidth]{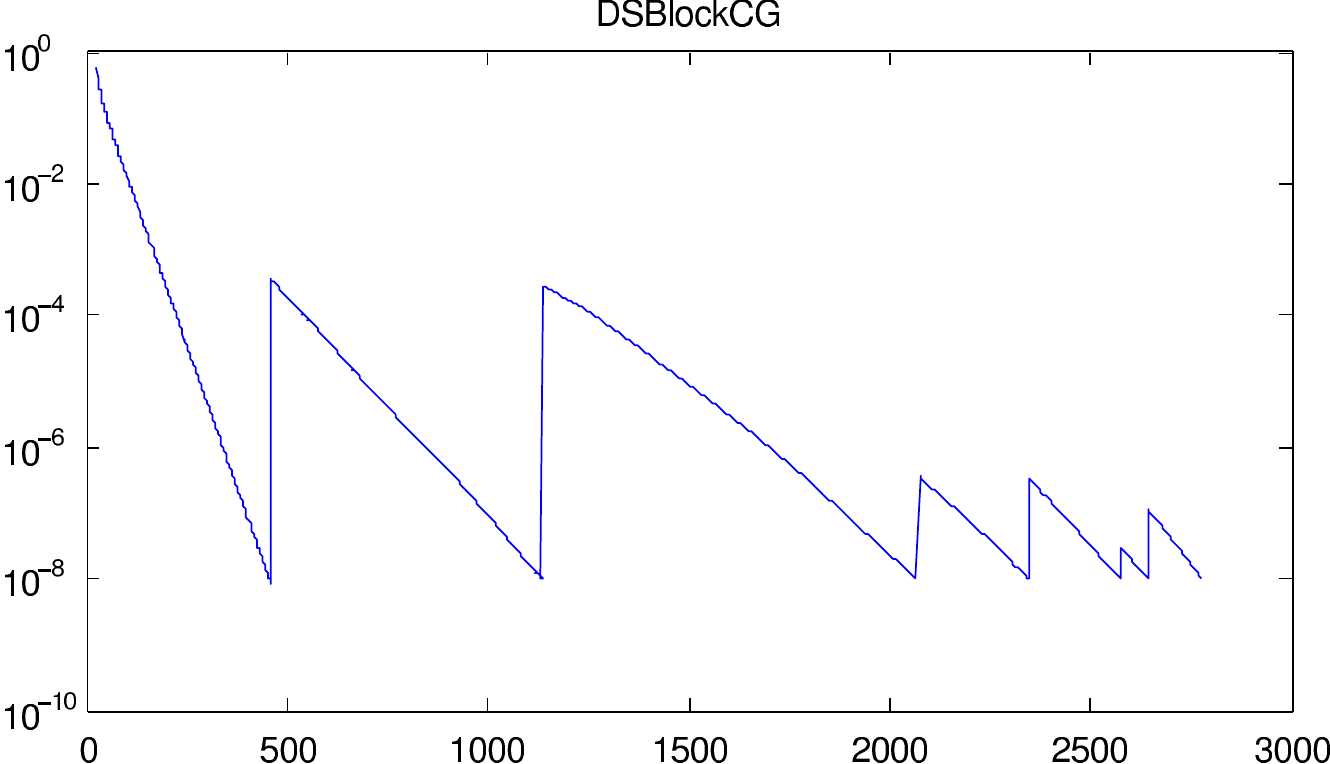}
  \hfill \includegraphics[width=0.48\textwidth]{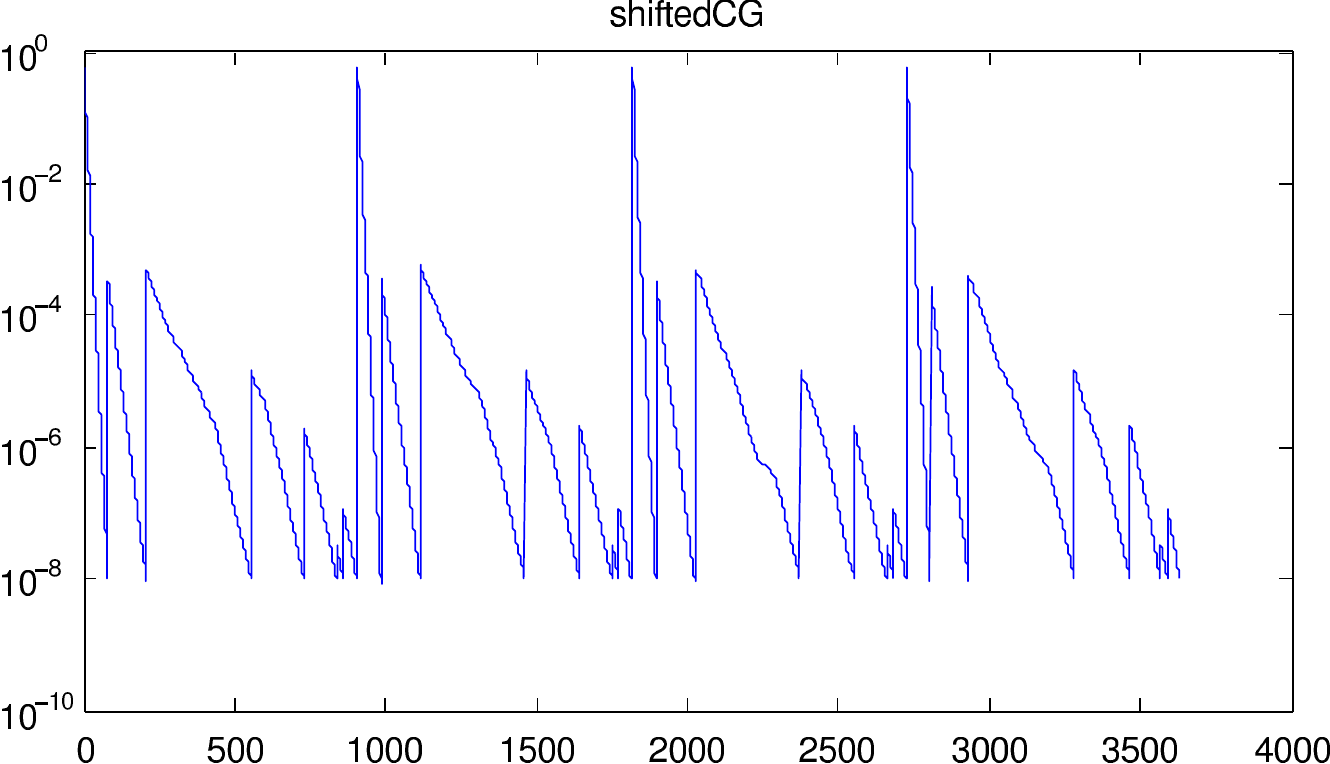}}
  \vspace{2ex}
  \centerline{\includegraphics[width=0.48\textwidth]{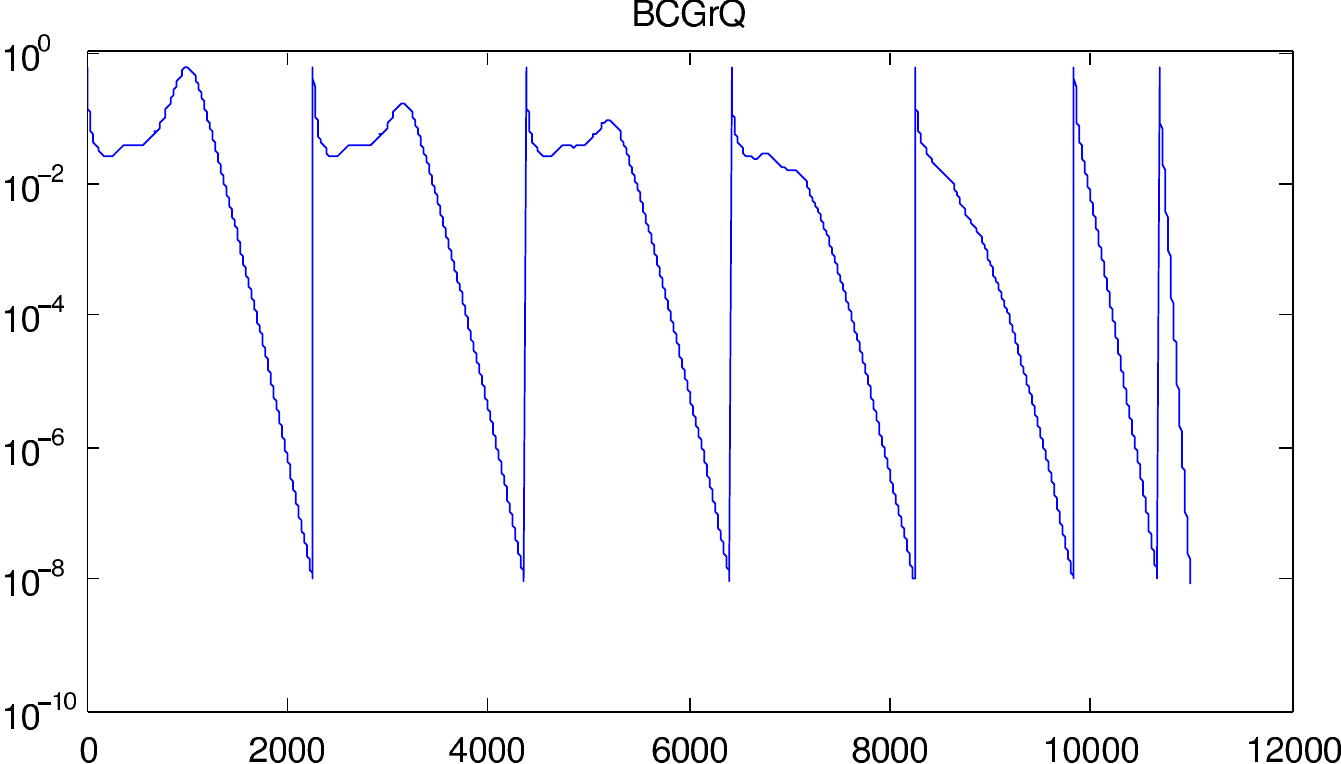}
  \hfill \includegraphics[width=0.48\textwidth]{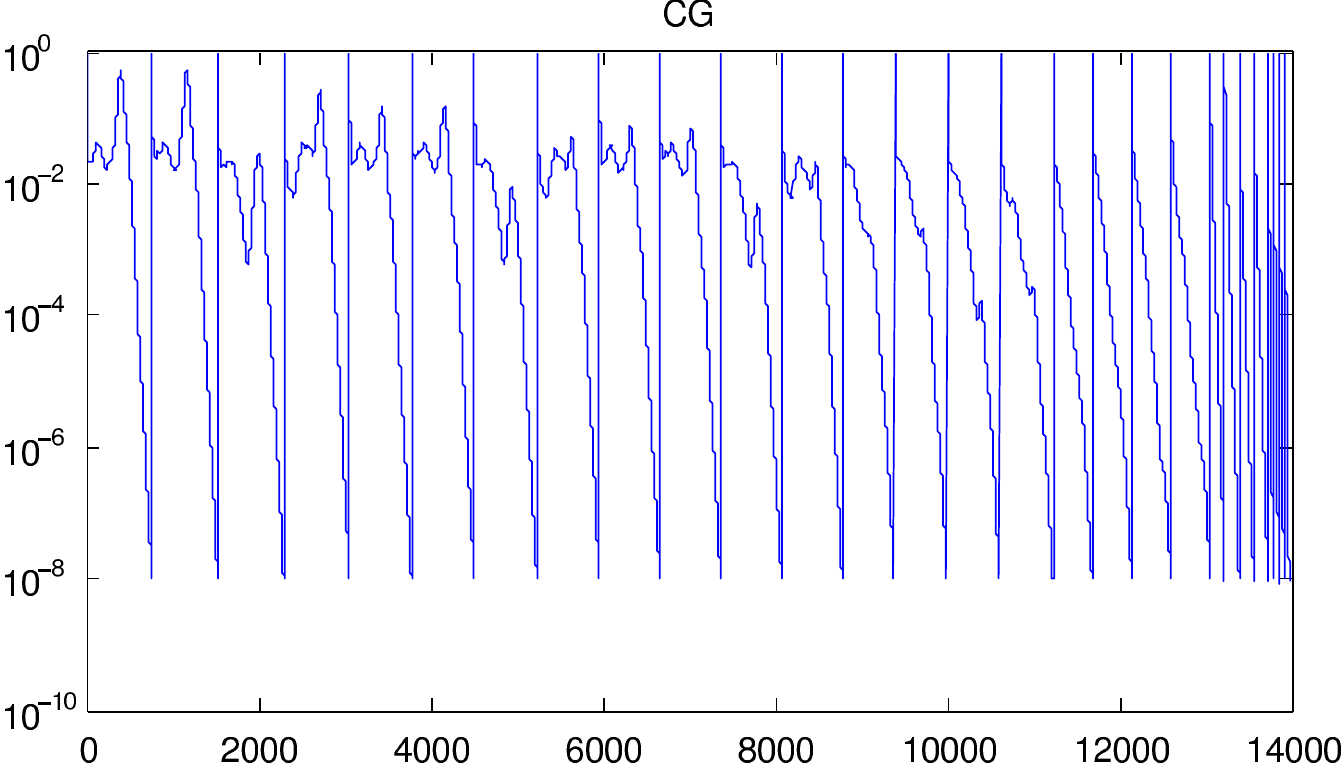}}
  \caption{Convergence plots for all the compared algorithms on a $16^4$ lattice with 4 \rhss{} and 7 shifts. Horizontal axis: time in seconds. Vertical axis: relative norm of residual. The saw tooth shape of the plots has two causes: a) for DSBlockCG and shiftedCG only the residual for the worst conditioned non-converged system is computed and plotted. b) methods that can not solve all of the systems at the same time need subsequent runs which are plotted one after the other. \label{fig_convergence}}
\end{figure}
Figure \ref{fig_convergence} shows a generic run of all four algorithms for a $16^4$ lattice to a target relative residual of $10^{-8}$. We chose the same $4$ random \rhss{} for all methods. The $7$ shifts were chosen, s.t. the smallest eigenvalue of the worst conditioned shifted system was of magnitude $10^{-6}$ and for the best conditioned system it was of magnitude $10^{-1}$.
DSBlockCG in the top left plot shows a saw tooth like convergence. This is caused by just computing the residual for the best conditioned not yet converged system which saves some work. Every time the plot of the relative residual jumps up, all the shifted systems for one \rhs{} have converged and the residuals for next non-converged system are computed. Thus, this plot shows that it is important to stop updating the iterates of well conditioned systems as soon as they reach the target residual. This speeds up the computation of the remaining systems.
The top right shiftedCG plot shows that the different random \rhs{} vectors show almost the same convergence behaviour. The bottom two plots show clearly that CG and BCGrQ are not competitive. Table \ref{tab_convergence} gives the plain numbers for the same test run and shows that shiftedCG takes about $30\%$ more time than DSBlockCG. The gap in the number of matrix-vector multiplications (mvms) is even bigger. As a result of Figure \ref{fig_convergence} we can constrain the remaining tests to comparisons of DSBlockCG and shiftedCG.

\begin{table}[ht]
  \begin{center}
    \begin{tabular}{|r|rrrr|}\hline
                        & CG      & BlockCG & shifted CG & DSBlockCG\\\hline
      mvms              & 22337   & 13488   & 4883       & 2766     \\
      time in seconds   & 13970.7 & 10993.3 & 3630.53    & 2778.82  \\
      relative time     & 5.03    & 3.96    & 1.31       & 1        \\\hline
    \end{tabular}
  \end{center}
  \caption{Number of mvms, time and relative time as compared to DSBlockCG. \label{tab_convergence}}
\end{table}

\begin{figure}[ht]
  \centerline{\includegraphics[width=0.48\textwidth]{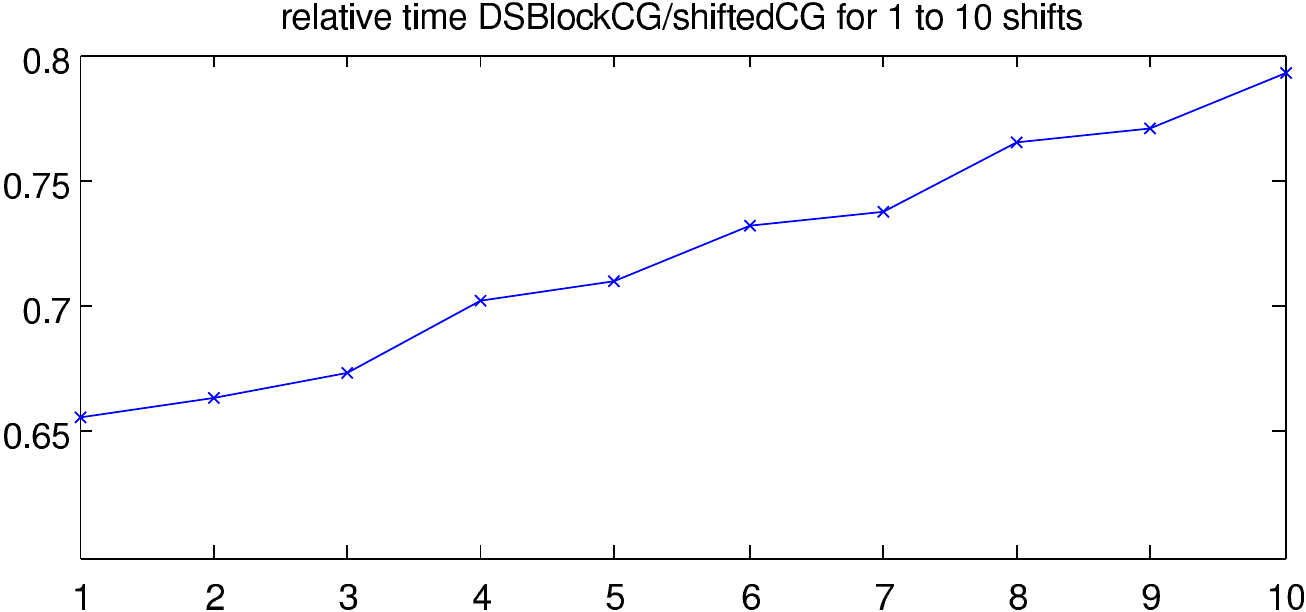}
  \hfill \includegraphics[width=0.48\textwidth]{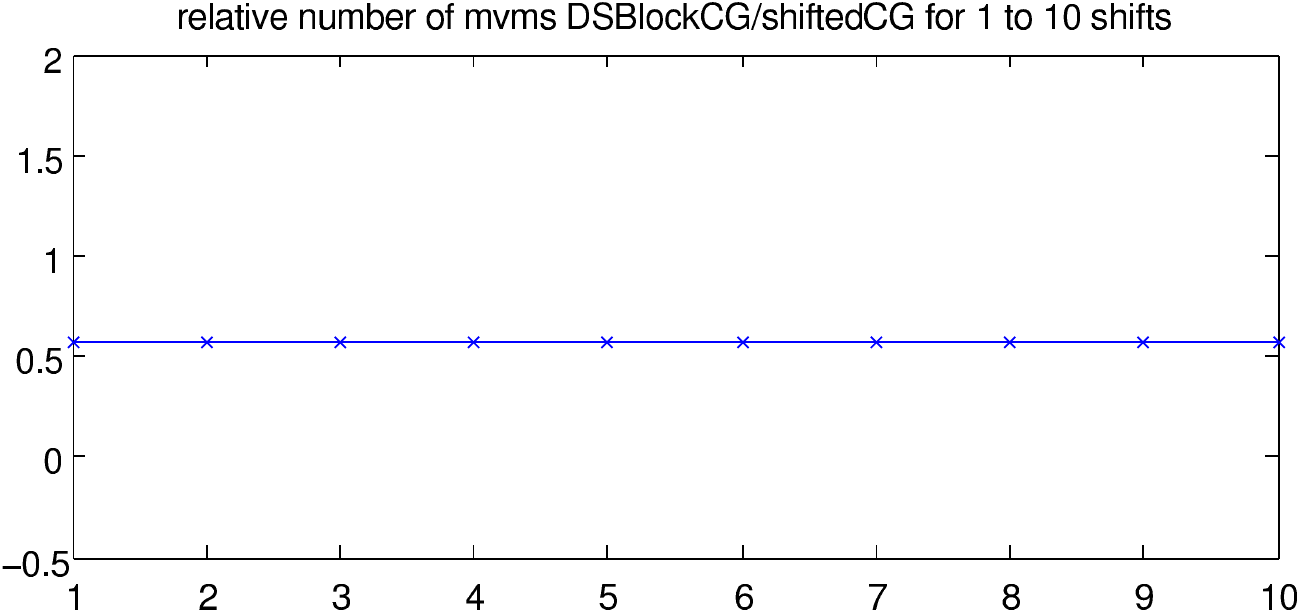}}
  \caption{The left plot shows the relative time (vertical axis) of algorithm DSBlockCG over shiftedCG on a $16^4$ lattice for a different number of shifts (horizontal axis) and a fixed number of 4 rhs. The right plot displays the relative number of mvms for the same run.\label{fig_shifts}}
\end{figure}
Figure \ref{fig_shifts} displays the dependence on the number of shifts while the number of \rhss{} stayed fixed at 4. The results were produced using the same $16^4$ lattice configuration as in the previous example. We chose the first shift, s.t. the smallest eigenvalue of the shifted system was of magnitude $10^{-6}$. We then successively increased the number of shifts, s.t. the smallest eigenvalues of the shifted systems were distributed in the interval $[10^{-6},10^{-1}]$. The plots show that the number of mvms depends only on the worst conditioned system which means that for any number of shifts DSBlockCG only needs about half the number of mvms compared to shiftedCG. The computational costs on the other hand are rising for an increased number of shifts.

\begin{figure}[ht]
  \centerline{\includegraphics[width=0.48\textwidth]{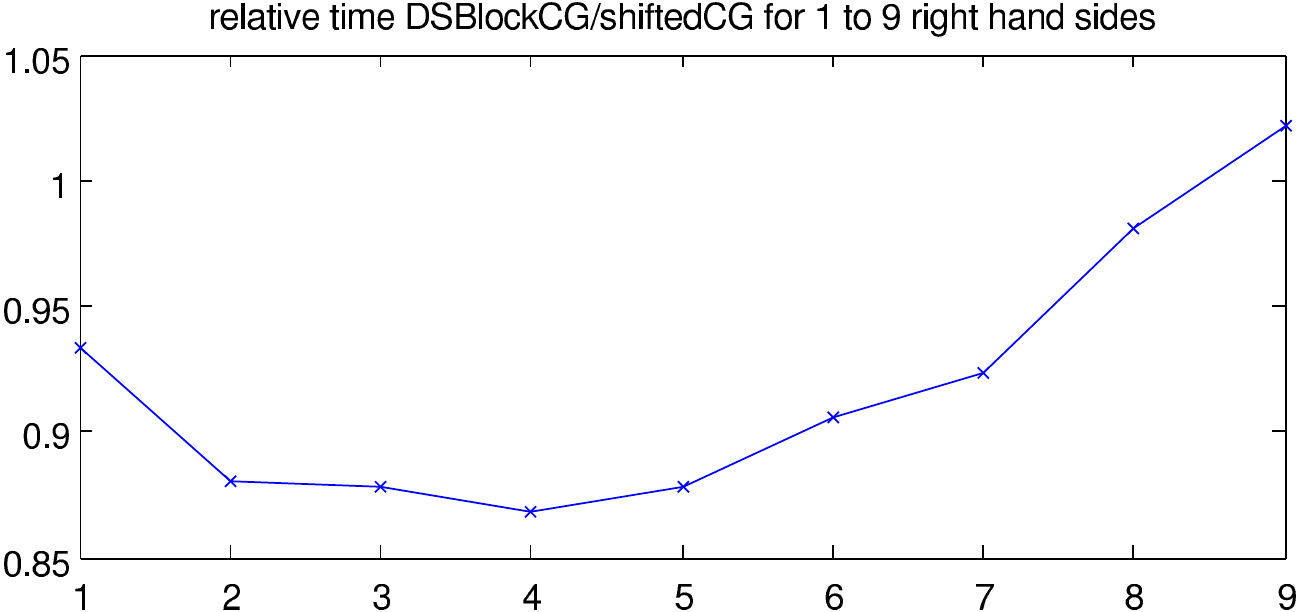}
  \hfill \includegraphics[width=0.48\textwidth]{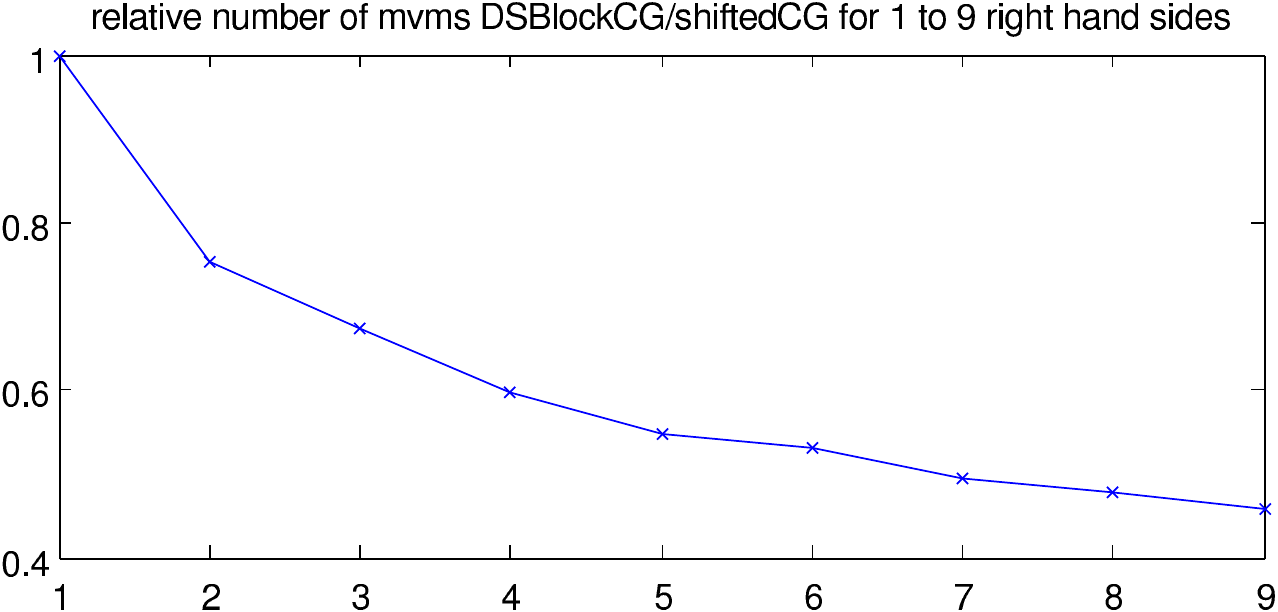}}
  \caption{The left plot shows the relative time (vertical axis) of algorithm DSBlockCG over shiftedCG on a $12^4$ lattice for a different number of \rhss{} (horizontal axis) and fixed number of 10 shifts. The right plot displays the relative number of mvms for the same run.\label{fig_rhs}}
\end{figure}
In Figure \ref{fig_rhs} we display a test run for another test configuration on a $12^4$ lattice. In this case the number of 10 shifts stayed fixed. Like in the previous test cases the shifts were chosen, s.t. the smallest eigenvalue of the worst conditioned shifted system was of magnitude $10^{-6}$ and for the best conditioned system it was of magnitude $10^{-1}$. There is a sweet spot at 4 shifts after which the additional costs start to become dominant and DSBlockCG becomes slower than shiftedCG, eventually.

\section{Conclusions}\label{sec_Conclusions}

We proposed a new iterative method for the solution of systems of linear equations of the form $(\sigma_jI+A)x_{i,j} = b_k$ based on a block Lanczos-type process. This method was shown to converge faster than just applying CG to every system or applying block CG methods to systems belonging to a single shift. For reasonable numbers of \rhss{} and shifts our new method even proved to be faster that shiftedCG.

\vfill

\bibliographystyle{abbrv}
\bibliography{./literature}

\begin{thebibliography}{10}

\bibitem{AliagaBoleyFreundHernandez2000}
J.~I. Aliaga, D.~L. Boley, R.~Freund, and V.~Hern{\'a}ndez.
\newblock A {L}anczos-type method for multiple starting vectors.
\newblock {\em Math. Comp.}, 69(232):1577--1601, 2000.

\bibitem{BirkFrommer2011DSBlockCG}
S.~Birk and A.~Frommer.
\newblock A deflated conjugate gradient method for multiple right hand sides
  and multiple shifts.
\newblock In preparation.

\bibitem{boost:lib}
Boost.
\newblock {B}oost {C++} libraries.
\newblock http://www.boost.org, 2010.
\newblock Version 1.44.0.

\bibitem{Dubrulle2001}
A.~A. Dubrulle.
\newblock Retooling the method of block conjugate gradients.
\newblock {\em Electron. Trans. Numer. Anal.}, 12:216--233 (electronic), 2001.

\bibitem{FreundFrommerFiebach1997}
P.~Fiebach, A.~Frommer, and R.~Freund.
\newblock Variants of the {B}lock-{QMR} method and applications in quantum
  chromodynamics.
\newblock In {\em 15th IMACS World Congress on Scientific Computation,
  Modelling and Applied Mathematics}, volume~3, pages 491--496, 1997.

\bibitem{FrommerMaass1999}
A.~Frommer and P.~Maass.
\newblock Fast {CG}-based methods for {T}ikhonov-{P}hillips regularization.
\newblock {\em SIAM J. Sci. Comput.}, 20(5):1831--1850 (electronic), 1999.

\bibitem{Gutknecht2005}
M.~H. Gutknecht.
\newblock {\em Modern Mathematical Models, Methods and Algorithms for Real
  World Systems}, chapter Block Krylov Space Methods for Linear Systems With
  Multiple Right-hand Sides: an Introduction, pages 420--447.
\newblock Anamaya Publishers, New Delhi, India, 2007.

\bibitem{Kennedy2006}
A.~D. Kennedy.
\newblock Algorithms for dynamical fermions.
\newblock Technical report, School of Physics, University of Edinburgh, 2006.
\newblock arXiv:hep-lat/0607038v1.

\bibitem{O'Leary1980}
D.~P. O'Leary.
\newblock The block conjugate gradient algorithm and related methods.
\newblock {\em Linear Algebra Appl.}, 29:293 -- 322, 1980.

\bibitem{paige1995approximate}
C.~C. Paige, B.~N. Parlett, and H.~A. van~der Vorst.
\newblock Approximate solutions and eigenvalue bounds from {K}rylov subspaces.
\newblock {\em Numer. Linear Algebra Appl.}, 2(2):115--133, 1995.

\end{thebibliography}

\end{document}